\documentclass[aps,prb,citeautoscript,twocolumn,nobibnotes,superscriptaddress,nofootinbib,10pt]{revtex4-1} 
\usepackage[latin1]{inputenc}
\usepackage{amsmath,amssymb}
\usepackage{mathrsfs} 
\usepackage{graphicx,color,colortbl}
\usepackage{rotating,array,tabularx,booktabs}
\usepackage{varwidth,xcolor}
\usepackage{placeins}
\usepackage{siunitx}

\DeclareGraphicsExtensions{.pdf,.PDF,.png,.PNG}

\newcommand{\dif}{\mathrm{d}}%
\newcommand{\ZT}[1]{\textquotedblleft#1\textquotedblright}%

\begin{document}

\title{Acoustically propelled nano- and microcones: fast forward and backward motion}

\author{Johannes Vo\ss{}}
\affiliation{Institut f\"ur Theoretische Physik, Center for Soft Nanoscience, Westf\"alische Wilhelms-Universit\"at M\"unster, D-48149 M\"unster, Germany}

\author{Raphael Wittkowski}
\email[Corresponding author: ]{raphael.wittkowski@uni-muenster.de}
\affiliation{Institut f\"ur Theoretische Physik, Center for Soft Nanoscience, Westf\"alische Wilhelms-Universit\"at M\"unster, D-48149 M\"unster, Germany}

\begin{abstract}
We focus on cone-shaped nano- and microparticles, which have recently been found to show particularly strong propulsion when they are exposed to a traveling ultrasound wave, and study based on direct acoustofluidic computer simulations how their propulsion depends on the cones' aspect ratio. The simulations reveal that the propulsion velocity and even its sign are very sensitive to the aspect ratio, where short particles move forward whereas elongated particles move backward. Furthermore, we identify a cone shape that allows for a particularly large propulsion speed. Our results contribute to the understanding of the propulsion of ultrasound-propelled colloidal particles, suggest a method for separation and sorting of nano- and microcones concerning their aspect ratio, and provide useful guidance for future experiments and applications. 
\begin{figure}[htb]
\centering
\fbox{\includegraphics[width=8cm]{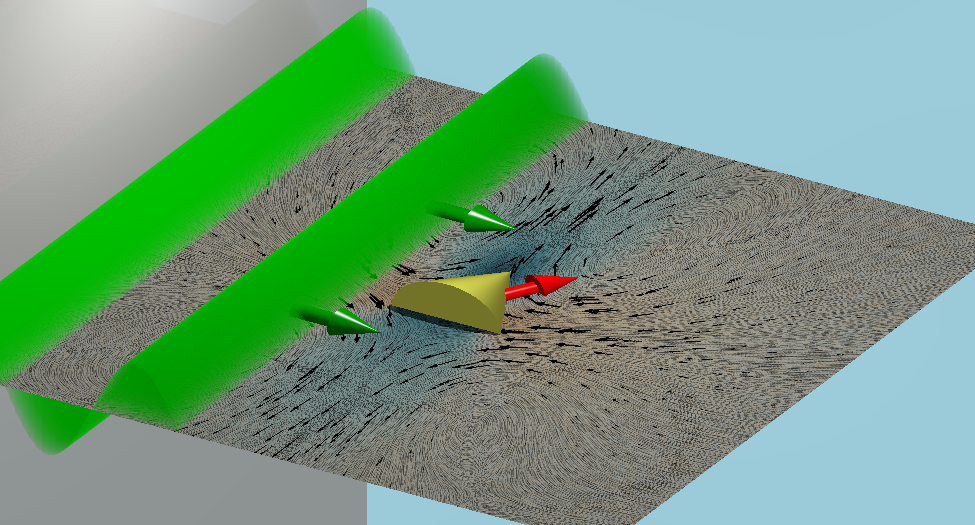}}%
\end{figure}
\end{abstract}
\maketitle

\section{Introduction}
Ultrasound-propelled colloidal particles, having been discovered in experiments in 2012 \cite{WangCHM2012}, constitute a particularly advantageous type of active nano- and microparticles \cite{BechingerdLLRVV2016,Venugopalan2020,FernandezRMHSS2020,YangEtAl2020}. The most important advantages of motile particles that are propelled by ultrasound   \cite{WangCHM2012,GarciaGradillaEtAl2013,AhmedEtAl2013,WuEtAl2014,WangLMAHM2014,GarciaGradillaSSKYWGW2014,BalkEtAl2014,AhmedGFM2014,EstebanFernandezdeAvilaMSLRCVMGZW2015,WuEtAl2015a,WuEtAl2015b,EstebanEtAl2016,SotoWGGGLKACW2016,AhmedWBGHM2016,AhmedBJPDN2016,UygunEtAl2017,EstebanFernandezEtAl2017,HansenEtAl2018,SabrinaTABdlCMB2018,WangGWSGXH2018,EstebanEtAl2018,LuSZWPL2019,QualliotineEtAl2019,GaoLWWXH2019,RenEtAl2019,VossW2020,RaoLMZCW2015,KaynakONLCH2017,ZhouYWDW2017,ZhouZWW2017,RenWM2018,ValdezLOESSWG2020,AghakhaniYWS2020,LiuR2020,DumyJMBGMHA2020} compared to particles with other propulsion mechanisms \cite{EstebanFernandezdeAvilaALGZW2018,SafdarKJ2018,PengTW2017,KaganBCCEEW2012,XuanSGWDH2018,XuCLFPLK2019} are the fact that the former particles can move in various types of fluids and soft materials, the bio-compatibility of their propulsion mechanism, and the easy way of supplying the particles permanently with energy. As a consequence, ultrasound-propelled nano- and microparticles have a number of important potential applications \cite{XuXZ2017}.   
An example is their usage as self-propelled nano- or microdevices in medicine \cite{LiEFdAGZW2017,PengTW2017,SotoC2018,WangGZLH2020,WangZ2021}, e.g., for targeted drug delivery \cite{LuoFWG2018,ErkocYCYAS2019}. 
There exist two different types of acoustically propelled particles: rigid particles \cite{WangCHM2012,GarciaGradillaEtAl2013,AhmedEtAl2013,NadalL2014,BalkEtAl2014,AhmedGFM2014,GarciaGradillaSSKYWGW2014,WangDZSSM2015,EstebanFernandezdeAvilaMSLRCVMGZW2015,SotoWGGGLKACW2016,EstebanEtAl2016,AhmedBJPDN2016,UygunEtAl2017,CollisCS2017,HansenEtAl2018,EstebanEtAl2018,SabrinaTABdlCMB2018,TangEtAl2019,ZhouZWW2017,VossW2020,ValdezLOESSWG2020,Zhou2018,RenWM2018,DumyJMBGMHA2020} and particles with movable components \cite{KaganBCCEEW2012,AhmedLNLSMCH2015,AhmedBJPDN2016,KaynakONLCH2017,ZhouYWDW2017,WangGWSGXH2018,RenEtAl2019,AghakhaniYWS2020,LiuR2020}. The rigid particles are easier to produce in large numbers and thus of special relevance with respect to future applications, where usually a large number of particles is required. There exist also some hybrid particles that combine acoustic propulsion with other propulsion mechanisms \cite{LILXKLWW2015,WangDZSSM2015,RenZMXHM2017,Zhou2018,TangEtAl2019,ValdezLOESSWG2020}.

In recent years, ultrasound-propelled nano- and microparticles have been intensively investigated \cite{WangCHM2012,GarciaGradillaEtAl2013,AhmedEtAl2013,BalkEtAl2014,NadalL2014,WangLMAHM2014,EstebanFernandezdeAvilaMSLRCVMGZW2015,SotoWGGGLKACW2016,EstebanEtAl2016,KaynakONNLCH2016,AhmedWBGHM2016,AhmedBJPDN2016,KaynakONLCH2017,CollisCS2017,HansenEtAl2018,SabrinaTABdlCMB2018,EstebanEtAl2018,TangEtAl2019,RaoLMZCW2015,KimGLZF2016,ZhouZWW2017,ChenEtAl2018,WangGWSGXH2018,LiuR2020,ZhouYWDW2017,Zhou2018,VossW2020,ValdezLOESSWG2020,AghakhaniYWS2020,LiuR2020,DumyJMBGMHA2020}. Besides only two articles that are based on analytical approaches \cite{NadalL2014,CollisCS2017} and an article that relies on direct computational fluid dynamics simulations \cite{VossW2020}, a large number of experimental studies have been published so far \cite{WangCHM2012,GarciaGradillaEtAl2013,AhmedEtAl2013,WangLMAHM2014,BalkEtAl2014,SotoWGGGLKACW2016,KaynakONNLCH2016,AhmedWBGHM2016,AhmedBJPDN2016,KaynakONLCH2017,SabrinaTABdlCMB2018,WangGWSGXH2018,TangEtAl2019,ZhouZWW2017,GaoLWWXH2019,RenEtAl2019,ZhouYWDW2017,Zhou2018,ValdezLOESSWG2020,DumyJMBGMHA2020}. 
In the previous work, mostly cylindrical particles with a concave end and a convex end were studied \cite{WangCHM2012,GarciaGradillaEtAl2013,AhmedEtAl2013,AhmedGFM2014,BalkEtAl2014,WangLMAHM2014,EstebanFernandezdeAvilaMSLRCVMGZW2015,AhmedWBGHM2016,ZhouZWW2017,WangGWSGXH2018,ZhouYWDW2017,Zhou2018,DumyJMBGMHA2020}. As a limiting case, which corresponds to a very short cylindrical particle with concave and convex ends, also half-sphere cups (nanoshells) were considered \cite{SotoWGGGLKACW2016,TangEtAl2019,VossW2020}. Recently, half-sphere-shaped particles, cone-shaped particles, and spherical- as well as conical-cup-like particles were compared with respect to their propulsion \cite{VossW2020}.   
In two other studies, gear-shaped microspinners \cite{KaynakONNLCH2016,SabrinaTABdlCMB2018} were addressed, and there are a few additional publications that focus on particles with movable components \cite{KaganBCCEEW2012,AhmedLNLSMCH2015,KaynakONLCH2017,AhmedBJPDN2016,RenEtAl2019}. 

Among the rigid particles with mainly translational motion that have been addressed so far, cone-shaped particles and conical-cup-shaped particles showed the fastest propulsion, where the speeds of cone-shaped and conical-cup-shaped particles differed only slightly \cite{VossW2020}. Since cone-shaped particles have a simpler shape, which facilitates their fabrication, and a larger volume, which is advantageous for delivery of drugs or other substances, than conical-cup-shaped particles, the former particles have been identified as particularly suitable candidates for efficient ultrasound-propelled particles that could be used in future experiments and applications.       
Cone-shaped particles can be produced, e.g., by electrodeposition \cite{LiRW2016,EstebanFernandezEtAl2017}, or directly be found, e.g., in the form of carbon nanocones \cite{TsakadzeLOX2007}, in large numbers.  
Up to now, however, only ultrasound-propelled cone-shaped particles with a particular aspect ratio have been studied \cite{VossW2020}, although the aspect ratio can have a strong influence on the efficiency of the particles' propulsion \cite{CollisCS2017}. 
Given that previous studies found for the short spherical-cup-shaped particles motion towards the particles' convex end \cite{SotoWGGGLKACW2016,TangEtAl2019,VossW2020} but for the longer cylindrical particles with concave and convex ends motion towards the concave end \cite{AhmedWBGHM2016}, also the direction of propulsion can depend on the aspect ratio. 

Therefore, in this article we investigate the acoustic propulsion of cone-shaped nano- and microparticles in more detail. Using direct acoustofluidic simulations, we study how the propulsion of acoustically propelled nano- and microcones depends on their aspect ratio and we determine an aspect ratio that is associated with particularly fast and thus efficient propulsion.

\section{\label{methods}Methods}
This work is based on a similar setup and procedure as Ref.\ \onlinecite{VossW2020}. We consider a particle that is surrounded by water and exposed to ultrasound. Using direct acoustofluidic simulations, where the compressible Navier-Stokes equations are solved numerically, the propagation of ultrasound through the water and the interaction with the particle are calculated. These calculations allow to determine the sound-induced forces and torques acting on the particle, from which in turn we calculate the particle's translational and angular propulsion velocities. 

\begin{figure}[htb]
\centering
\includegraphics[width=\linewidth]{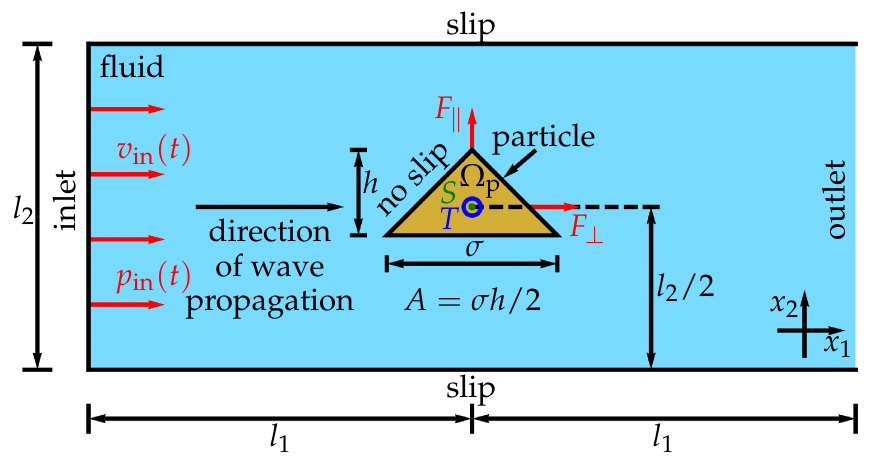}%
\caption{\label{fig:setup}The considered setup. A rigid cone-shaped particle is in the middle of a fluid-filled rectangular domain. The particle has width $\sigma$, height $h$, and a fixed cross-section area $A$, and is described by a particle domain $\Omega_{\mathrm{p}}$. Furthermore, the rectangular domain has width $2 l_1$ and height $l_2$ and the center of mass $\mathrm{S}$ of the particle is in the middle of the rectangular domain. At the inlet, a traveling ultrasound wave is entering the fluid-filled domain. For this purpose, an inflow velocity $v_{\mathrm{in}}(t)$ and pressure $p_{\mathrm{in}}(t)$ are prescribed at the inlet. The ultrasound wave propagates through the fluid, where slip boundary conditions are used for the lateral boundaries of the rectangular domain. At the particle, for which no-slip boundary conditions are used, the ultrasound exerts a propulsion force with time-averaged components $F_{\parallel}$ and $F_\perp$ parallel and perpendicular to the particle's orientation, respectively, as well as a time-averaged torque $T$. When the ultrasound wave reaches the end of the domain, it leaves the domain through the outlet.}
\end{figure}

Figure \ref{fig:setup} shows the setup in detail.
We consider a particle with a conical shape in two spatial dimensions. The particle is oriented perpendicular to the direction of wave propagation, has a fixed cross-section area $A$, and is described by a particle domain $\Omega_{\mathrm{p}}$. 
Its aspect ratio $\chi=h/\sigma$ with the particle's height $h$ and diameter $\sigma$ is varied. 
The position of the particle is fixed. This means that the results of the simulations are valid for a particle which is held in place. Such a particle can be seen as a free moving particle in the limiting case of an infinite mass density. This limiting case can, in turn, be considered as an upper bound for a free moving particle made of a material with a high mass density like gold, which is a widely used material for such particles \cite{WangCHM2012,WangLMAHM2014,BalkEtAl2014,EstebanFernandezdeAvilaMSLRCVMGZW2015,SotoWGGGLKACW2016,AhmedWBGHM2016,EstebanEtAl2016,ZhouZWW2017,EstebanFernandezEtAl2017,EstebanEtAl2018,HansenEtAl2018,SabrinaTABdlCMB2018,QualliotineEtAl2019}.

The particle is positioned in the middle of a water-filled rectangular domain so that the center of mass of the rectangle and the center of mass $\mathrm{S}$ of the particle coincide. 
One edge of the rectangular domain has length $l_2=\SI{200}{\micro\metre}$ and is perpendicular to the direction of ultrasound propagation. 
The other edge is parallel to the direction of ultrasound propagation and has length $2l_1$. 
We choose a Cartesian coordinate system such that the $x_1$ axis is parallel to the direction of ultrasound propagation and the $x_2$ axis is perpendicular to that direction, i.e., the coordinate axes are parallel to the edges of the rectangular domain. 
The ultrasound wave has frequency $f=\SI{1}{\MHz}$ and enters the rectangular domain at the edge perpendicular to the sound-propagation direction. 
We prescribe the incoming ultrasound wave by a time-dependent inflow pressure $p_{\mathrm{in}}(t)=\Delta p \sin(2\pi f t)$ and velocity $v_{\mathrm{in}}(t)=(\Delta p / (\rho_0 c_{\mathrm{f}})) \sin(2\pi f t)$ perpendicular to the inlet with the pressure amplitude $\Delta p=\SI{10}{\kilo \pascal}$, density of the initially quiescent fluid $\rho_0=\SI{998}{\kilogram\,\metre^{-3}}$, and sound velocity in the fluid $c_{\mathrm{f}}=\SI{1484}{\metre\,\second^{-1}}$. 

The acoustic energy density from this wave is
$E=\Delta p^2/(2 \rho_0 c_{\mathrm{f}}^2)=\SI{22.7}{\milli\joule\,\metre^{-3}}$.
After a distance $l_1=\lambda/4$, where $\lambda$ is the wavelength of the ultrasound wave $\lambda=\SI{1.484}{\milli\metre}$, the wave reaches the fixed rigid particle. 
The interaction of the ultrasound with the particle leads to time-averaged forces $F_\parallel$ parallel and $F_\perp$ perpendicular to the particle orientation as well as to a time-averaged torque $T$ relative to the reference point $\mathrm{S}$ acting on the particle.     
After a further distance $l_1$ the wave leaves the domain through an outlet at the edge of the water domain opposing the inlet. 
The particle boundaries are described by a no-slip condition and at the edges of the water domain parallel to the direction of sound propagation we assume a slip condition.  

In the simulations, we numerically solve the continuity equation for the mass-density field of the fluid, the compressible Navier-Stokes equation, and a linear constitutive equation for the fluid's pressure field. Thus we are avoiding approximations like perturbation expansions that are used in most previous studies using analytical \cite{NadalL2014,CollisCS2017} or numerical \cite{AhmedBJPDN2016,SabrinaTABdlCMB2018,TangEtAl2019} methods. For solving these equations, we used the finite volume method implemented in the software package OpenFOAM \cite{WellerTJF1998}.
We applied a structured mixed rectangular-triangular mesh with about 300,000 cells, where the cell size $\Delta x$ is very small close to the particle, and larger far away from it. Concerning the time integration, an adaptive time-step method is used with a time-step size $\Delta t$ such that the Courant-Friedrichs-Lewy number 
\begin{align}
C = c_\mathrm{f} \frac{\Delta t}{\Delta x}
\end{align}
is smaller than one. 
To get sufficiently close to the stationary state, we simulated a time interval with a duration of $t_\mathrm{max}= 500\tau$ or more, where $\tau$ is the period of the ultrasound wave.  
An individual simulation required a computational expense of typically $36,000$ CPU core hours. The reason for this expense is the necessary fine discretization in space and time relative to the large spatial and temporal domains.

Through the simulations, we calculated the time-dependent force and torque acting on the particle in the laboratory frame. Since the particle has no-slip boundary conditions and is fixed in space, the fluid velocity is zero at the fluid-particle interface. So the force and torque can be calculated by the integral of the stress tensor $\Sigma$ over the particle surface. The force $\vec{F}^{(p)}+\vec{F}^{(v)}$ and torque $T^{(p)}+T^{(v)}$ consist of two components, namely a pressure component (superscript \ZT{$(p)$}) and a viscosity component (superscript \ZT{$(v)$}) with \cite{LandauL1987}
{\allowdisplaybreaks\begin{align}%
F^{(\alpha)}_{i} &=  \sum^{2}_{j=1} \int_{\partial\Omega_{\mathrm{p}}} \!\!\!\!\!\!\! \Sigma^{(\alpha)}_{ij}\,\dif A_{j}, 
\label{eq:F}\\%
T^{(\alpha)} &=  \sum^{2}_{j,k,l=1} \int_{\partial\Omega_{\mathrm{p}}} \!\!\!\!\!\!\! \epsilon_{3jk}(x_j-x_{\mathrm{p},j})\Sigma^{(\alpha)}_{kl}\,\dif A_{l} 
\label{eq:T}%
\end{align}}%
for $\alpha\in\{p,v\}$. Here, $\Sigma^{(p)}$ and $\Sigma^{(v)}$ are the pressure component and the viscous component of the stress tensor, respectively.
$\dif\vec{A}(\vec{x})=(\dif A_{1}(\vec{x}),\dif A_{2}(\vec{x}))^{\mathrm{T}}$ is the normal and outwards oriented surface element of $\partial\Omega_{\mathrm{p}}$ at position $\vec{x}$ when $\vec{x}\in\partial\Omega_{\mathrm{p}}$, $\epsilon_{ijk}$ the Levi-Civita symbol, and $\vec{x}_{\mathrm{p}}$ the position of $\mathrm{S}$.
To obtain the time-averaged stationary values, we locally averaged over one period and extrapolated towards $t \to \infty$ using the extrapolation procedure described in Ref.\ \onlinecite{VossW2020}. 

With this procedure, we get the force $\vec{F}=\vec{F}_p+\vec{F}_v$ with pressure component $\vec{F}_p=\langle\vec{F}^{(p)}\rangle$ and viscous component $\vec{F}_v=\langle\vec{F}^{(v)}\rangle$ as well as the torque $T=T_p + T_v$ with components $T_p=\langle T^{(p)}\rangle$ and $T_v=\langle T^{(v)}\rangle$ acting on the particle, where $\langle\cdot\rangle$ denotes the time average. 
To calculate the translational-angular velocity vector $\vec{\mathfrak{v}}=(\vec{v},\omega)^{\mathrm{T}}$ with the particle's translational velocity $\vec{v}$ and angular velocity $\omega$, we define the force-torque vector $\vec{\mathfrak{F}}=(\vec{F},T)^{\mathrm{T}}$. Then the values of $\vec{\mathfrak{v}}$ can be calculated with the Stokes law as \cite{HappelB1991} 
\begin{equation}
\vec{\mathfrak{v}}=\frac{1}{\nu_\mathrm{s}} \boldsymbol{\mathrm{H}}^{-1}\vec{\mathfrak{F}} 
\label{eq:velocity}%
\end{equation}
with the fluid's shear viscosity $\nu_{\mathrm{s}}$ and the hydrodynamic resistance matrix
\begin{equation}
\boldsymbol{\mathrm{H}}=
\begin{pmatrix}
\boldsymbol{\mathrm{K}} & \boldsymbol{\mathrm{C}}^{\mathrm{T}}_{\mathrm{S}} \\
\boldsymbol{\mathrm{C}}_{\mathrm{S}} & \boldsymbol{\Omega}_{\mathrm{S}} 
\end{pmatrix} .
\label{eq:H}%
\end{equation}
Here, $\boldsymbol{\mathrm{K}}_{\mathrm{S}}$, $\boldsymbol{\mathrm{C}}_{\mathrm{S}}$, and $\boldsymbol{\Omega}_{\mathrm{S}}$ are $3\times 3$-dimensional submatrices and the subscript $\mathrm{S}$ denotes the particle's center of mass as the reference point. 

Since we consider a system with two spatial dimensions to keep the computational effort for the simulations manageable, we cannot use Eqs.\ \eqref{eq:F}-\eqref{eq:H} directly. Hence, we assign a thickness of $\SI{1}{\micro\metre}$ to the particle, so that $\boldsymbol{\mathrm{H}}$ can be calculated. The general structure of $\boldsymbol{\mathrm{H}}$ for a particle with a shape as we study here is
\begin{align}
\boldsymbol{\mathrm{H}} = \begin{pmatrix}
\mathrm{K_{11}} & 0 & 0 & 0 & 0 & \mathrm{C_{31}}\\
 0 & \mathrm{K_{22}} & 0 & 0 & 0 & 0 \\
 0 & 0 & \mathrm{K_{33}} & \mathrm{C_{13}} & 0 & 0\\
 0 & 0 & \mathrm{C_{13}} & \mathrm{\Omega_{11}} & 0 & 0 \\
 0 & 0 & 0 & 0 & \mathrm{\Omega_{22}} & 0 \\
 \mathrm{C_{31}} & 0 & 0 & 0 & 0 &\mathrm{\Omega_{33}}
\end{pmatrix},
\end{align}
where the values of the nonzero elements are given in Tab.\ \ref{tab:resistance} for each aspect ratio considered in this work.
\begin{table*}[tb]
\centering
\begin{ruledtabular}
\begin{tabular}{ccccccccc}
$\boldsymbol{\chi}$ & $\mathbf{K_{11} / \SI{}{\textbf{\micro\metre}}}$ & $\mathbf{K_{22}} / \SI{}{\textbf{\micro\metre}}$ & $\mathbf{K_{33}} / \SI{}{\textbf{\micro\metre}}$ & $\mathbf{C_{13}} / \SI{}{\textbf{\micro\metre}^{\textbf{2}}}$ & $\mathbf{C_{31}} / \SI{}{\textbf{\micro\metre}^{\textbf{2}}}$& $\mathbf{\Omega_{11}} / \SI{}{\textbf{\micro\metre}^{\textbf{3}}}$ & $\mathbf{\Omega_{22}} / \SI{}{\textbf{\micro\metre}^{\textbf{3}}}$ & $\mathbf{\Omega_{33}} / \SI{}{\textbf{\micro\metre}^{\textbf{3}}}$ \\
\hline
$\SI{0.25}{}$ & $\SI{8.58}{}$ & $\SI{11.29}{}$ & $\SI{8.98}{}$ & $\SI{-0.38}{}$ & $\SI{0.61}{}$ & $\SI{3.44}{}$ & $\SI{4.76}{}$ & $\SI{4.75}{}$\\
$\SI{0.5}{}$ & $\SI{8.49}{}$ & $\SI{9.72}{}$ & $\SI{8.14}{}$ & $\SI{-0.14}{}$ & $\SI{0.4}{}$ & $\SI{3}{}$ & $\SI{3.36}{}$ & $\SI{2.63}{}$\\
$\SI{0.75}{}$ & $\SI{8.76}{}$ & $\SI{9.07}{}$ & $\SI{7.94}{}$ & $\SI{-0.03}{}$ & $\SI{0.11}{}$ & $\SI{3.06}{}$ & $\SI{3.03}{}$ & $\SI{2.22}{}$\\
$\SI{1}{}$ & $\SI{9.03}{}$ & $\SI{8.77}{}$ & $\SI{7.94}{}$ & $\SI{-0.01}{}$ & $\SI{-0.14}{}$ & $\SI{3.18}{}$ & $\SI{2.96}{}$ & $\SI{2.27}{}$\\
$\SI{1.5}{}$ & $\SI{9.69}{}$ & $\SI{8.57}{}$ & $\SI{8.13}{}$ & $\SI{0.25}{}$ & $\SI{-0.54}{}$ & $\SI{3.68}{}$ & $\SI{2.97}{}$ & $\SI{2.68}{}$\\
$\SI{2}{}$ & $\SI{10.15}{}$ & $\SI{8.55}{}$ & $\SI{8.39}{}$ & $\SI{0.43}{}$ & $\SI{-0.9}{}$ & $\SI{4.25}{}$ & $\SI{3.08}{}$ & $\SI{3.29}{}$\\
$\SI{2.5}{}$ & $\SI{10.61}{}$ & $\SI{8.56}{}$ & $\SI{8.59}{}$ & $\SI{0.6}{}$ & $\SI{-1.21}{}$ & $\SI{4.86}{}$ & $\SI{3.17}{}$ & $\SI{3.97}{}$\\
$\SI{3}{}$ & $\SI{11.02}{}$ & $\SI{8.64}{}$ & $\SI{8.84}{}$ & $\SI{0.77}{}$ & $\SI{-1.5}{}$ & $\SI{5.53}{}$ & $\SI{3.29}{}$ & $\SI{4.69}{}$\\
$\SI{3.5}{}$ & $\SI{11.41}{}$ & $\SI{8.75}{}$ & $\SI{9.08}{}$ & $\SI{0.95}{}$ & $\SI{-1.79}{}$ & $\SI{6.23}{}$ & $\SI{3.41}{}$ & $\SI{5.47}{}$\\
$\SI{4}{}$ & $\SI{11.67}{}$ & $\SI{8.87}{}$ & $\SI{9.33}{}$ & $\SI{1.1}{}$ & $\SI{-2.07}{}$ & $\SI{6.94}{}$ & $\SI{3.51}{}$ & $\SI{6.27}{}$\\
\end{tabular} 
\end{ruledtabular}%
\caption{\label{tab:resistance}Elements of the hydrodynamic resistance matrix $\boldsymbol{\mathrm{H}}$ for a particle with a triangular cross section as shown in Fig.\ \ref{fig:setup}, a thickness of $\SI{1}{\micro\metre}$ in the third dimension, and the center of mass as the reference point for different aspect ratios $\chi=h/\sigma$ with the particle's height $h$ and diameter $\sigma$.}
\end{table*}
By neglecting the contributions $\mathrm{K_{33}}$, $\mathrm{C_{13}}$, $\mathrm{\Omega_{11}}$, and $\mathrm{\Omega_{22}}$ that correspond to the lower and upper surfaces of the particle, we can then use the three-dimensional versions of Eqs.\ \eqref{eq:F}-\eqref{eq:velocity}.

We determine the components of $\vec{F}$ and $\vec{v}$ parallel and perpendicular to the particle's orientation, i.e., parallel to the $x_2$ and $x_1$ axes, respectively.
These components are the parallel force $F_\parallel=(\vec{F})_2=F_{\parallel,p}+F_{\parallel,v}$, with its pressure component $F_{\parallel,p}=(\langle\vec{F}^{(p)}\rangle)_2$ and viscous component $F_{\parallel,v}=(\langle\vec{F}^{(v)}\rangle)_2$, perpendicular force $F_\perp = (\langle\vec{F}_\perp\rangle)_1=F_{\perp,p}+F_{\perp,v}$ with the components $F_{\perp,p}=(\langle\vec{F}^{(p)}\rangle)_1$ and $F_{\perp,v}=(\langle\vec{F}^{(v)}\rangle)_1$, parallel speed $v_{\parallel}=(\vec{v})_2$, and perpendicular speed $v_{\perp}=(\vec{v})_1$. 

Nondimensionalization of the governing equations leads to four dimensionless numbers: the Helmholtz number $\mathrm{He}$, a Reynolds number corresponding to the shear viscosity $\mathrm{Re}_\mathrm{s}$, another Reynolds number corresponding to the bulk viscosity $\mathrm{Re}_\mathrm{b}$, and the number $\mathrm{Ma}^2\mathrm{Eu}$ corresponding to the pressure amplitude $\Delta p$ of the ultrasound wave entering the simulated system, where $\mathrm{Ma}$ is the Mach number and $\mathrm{Eu}$ is the Euler number. Table \ref{tab:Parameters} shows the names and symbols of the parameters that are relevant for our simulations and their values that we have chosen in analogy to the values used in Ref.\ \onlinecite{VossW2020}.
\begin{table}[htb]
\centering
\begin{ruledtabular}
\begin{tabular}{p{46mm}ccc}%
\textbf{Name} & \textbf{Symbol} & \textbf{Value}\\
\hline
Particle cross-section area & $A$ & \SI{0.25}{\micro\metre^2}\\
Particle diameter-height ratio & $\chi=h/\sigma$ & $\SI{0.25}{}$-$\SI{4}{}$\\
Particle diameter & $\sigma$ & $\sqrt{2A/\chi}$\\
Particle height & $h$ & $\sigma\chi$\\
Sound frequency & $f$ & $\SI{1}{\mega\hertz}$\\
Speed of sound & $c_\mathrm{f}$ & $\SI{1484}{\metre\,\second^{-1}}$\\
Time period of sound & $\tau=1/f$ & $\SI{1}{\micro\second}$\\
Wavelength of sound & $\lambda=c_\mathrm{f}/f$ & $\SI{1.484}{\milli\metre}$\\
Temperature of fluid & $T_0$ & $\SI{293.15}{\kelvin}$\\
Mean mass density of fluid & $\rho_0$ & $\SI{998}{\kilogram\,\metre^{-3}}$\\
Mean pressure of fluid & $p_{0}$ & $\SI{101325}{\pascal}$ \\
Initial velocity of fluid & $\vec{v}_{0}$ & $\vec{0}\,\SI{}{\metre\,\second^{-1}}$ \\
Sound pressure amplitude & $\Delta p$ & \SI{10}{\kilo\pascal}\\
Acoustic energy density & \hspace*{-4mm}$E=\Delta p^2/(2 \rho_0 c_{\mathrm{f}}^2)$\hspace*{-1mm} & $\SI{22.7}{\milli\joule\,m^{-3}}$\\
Shear/dynamic viscosity of fluid & $\nu_{\mathrm{s}}$ & $\SI{1.002}{\milli\pascal\,\second}$ \\
Bulk/volume viscosity of fluid & $\nu_{\mathrm{b}}$ & $\SI{2.87}{\milli\pascal\,\second}$ \\
\mbox{Inlet-particle or particle-outlet} distance & $l_1$ & $\lambda/4$ \\
Inlet length & $l_2$ & $\SI{200}{\micro\metre}$\\
Mesh-cell size & $\Delta x$ & $\SI{15}{\nano \metre}$-$\SI{1}{\micro \metre}$ \\
Time-step size & $\Delta t$ & $1$-$\SI{10}{\pico \second}$\\
Simulation duration & $t_{\mathrm{max}}$ & $500\tau$ \\
\end{tabular}%
\end{ruledtabular}%
\caption{\label{tab:Parameters}Parameters that are relevant for our simulations and their values, which are chosen similar to those in Ref.\ \onlinecite{VossW2020}. The values of the speed of sound $c_\mathrm{f}$, mean mass density $\rho_0$, shear viscosity $\nu_\mathrm{s}$, and bulk viscosity $\nu_\mathrm{b}$ are calculated for water at rest at normal temperature $T_0$ and normal pressure $p_0$.}%
\end{table}
By using the parameter values from Tab.\ \ref{tab:Parameters}, the dimensionless numbers for our simulations have the following values:
\begin{align}
\mathrm{He}&=2 \pi f \sqrt{A} / c_\mathrm{f}\approx 2.117\cdot 10^{-3},\\
\mathrm{Re}_\mathrm{s}&=\rho_0 c_\mathrm{f} \sqrt{A} / \nu_\mathrm{s}\approx 739,\\
\mathrm{Re}_\mathrm{b}&=\rho_0 c_\mathrm{f} \sqrt{A} / \nu_\mathrm{b}\approx 258,\\
\mathrm{Ma}^2 \mathrm{Eu}&=\Delta p/(\rho_0 c_\mathrm{f}^2)\approx 4.550\cdot 10^{-6}.
\end{align}
Note that the Reynolds number $\mathrm{Re}=\rho_{0}\sqrt{A(v_\parallel^2+v_\perp^2)}/\nu_\mathrm{s}<6\cdot 10^{-8}$, which characterizes the particle motion through the fluid, is close to zero.

\section{\label{results}Results and discussion}
\begin{figure*}[p]
\centering
\includegraphics[width=\linewidth]{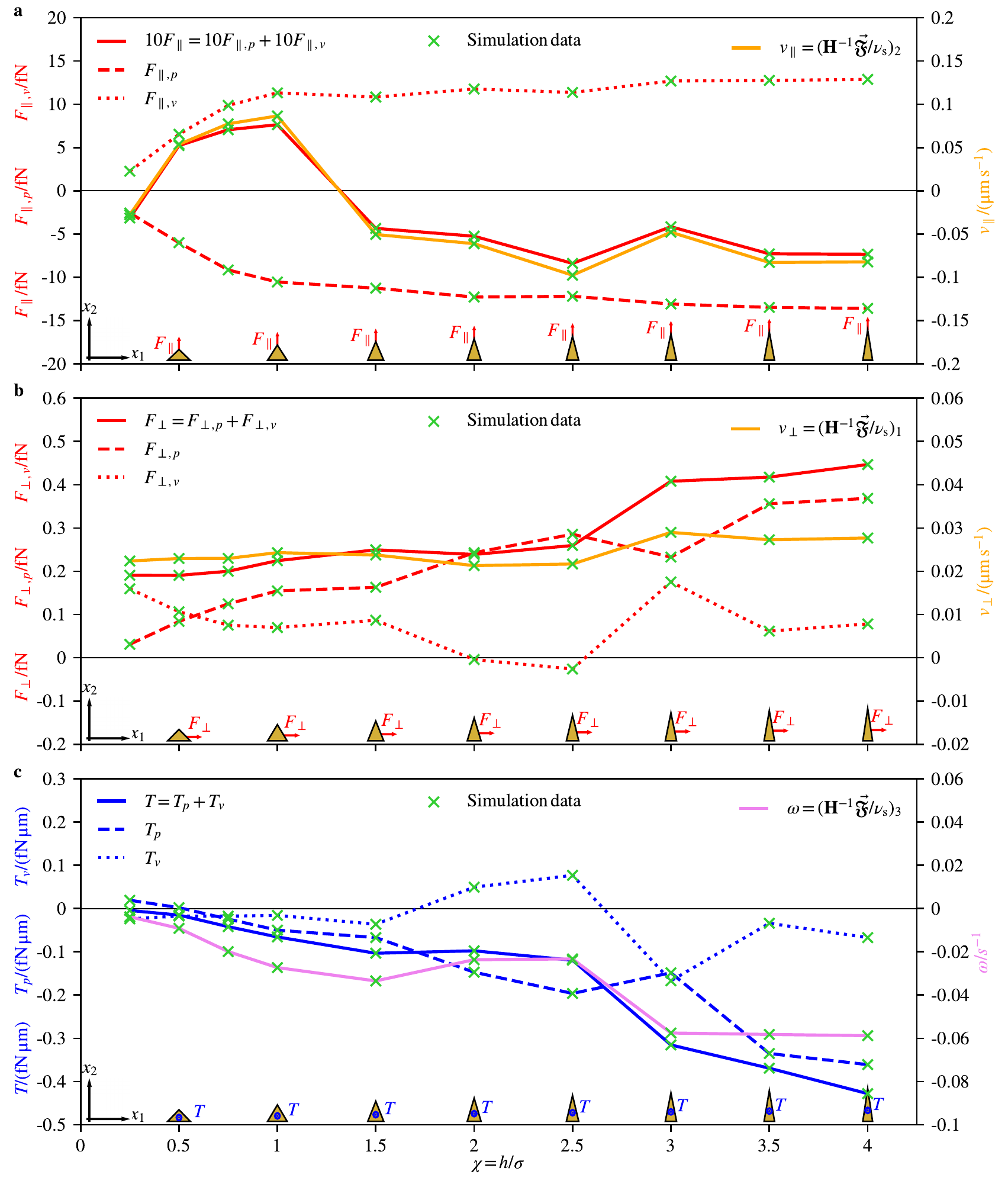}%
\caption{\label{fig:plot}(a) Simulation data for the forces $F_{\parallel,p}$ and $F_{\parallel,v}$ acting on a particle with triangular cross section and aspect ratio $\chi$ parallel to its orientation, their sum $F_\parallel=F_{\parallel,p}+F_{\parallel,v}$, and the corresponding propulsion velocity $v_\parallel$ for various values of $\chi$. (b) The corresponding forces $F_{\perp,p}$ and $F_{\perp,v}$ for the direction perpendicular to the particle orientation, their sum $F_\perp=F_{\perp,p} + F_{\perp,v}$, and the velocity $v_\perp$ for different values of $\chi$. (c) Results of the simulation for the torque components $T_p$ and $T_v$ acting on the particle, their sum $T=T_p+T_v$, and the corresponding angular velocity $\omega$ for different values of $\chi$.}
\end{figure*}

Figure \ref{fig:plot} shows our results for the propulsion-force components $F_\parallel$ and $F_\perp$, propulsion torque $T$, translational-propulsion-velocity components $v_\parallel$ and $v_\perp$, and angular propulsion velocity $\omega$ as functions of the aspect ratio $\chi\in\left[\SI{0.25}{},\SI{4}{}\right]$. 

The force $F_\parallel$ and velocity $v_\parallel$ have a strong dependence on the aspect ratio $\chi$. This includes even a sign change. Both curves have qualitatively the same course. They start at $\chi=\SI{0.25}{}$ with negative propulsion force $F_\parallel=\SI{-0.32}{\femto\newton}$ and velocity $v_\parallel=\SI{-0.028}{\micro\metre\,\second^{-1}}$. 
Increasing $\chi$ leads to a positive sign of $F_\parallel$ and $v_\parallel$ until about $\chi=\SI{1}{}$ where the force is maximal with  $F_\parallel=\SI{0.76}{\femto\newton}$ and also the speed reaches its maximum $v_\parallel=\SI{0.086}{\micro\metre\,\second^{-1}}$.
Afterwards, the propulsion force and velocity decrease to and remain at negative values. The globally maximal amplitude is reached at $\chi=\SI{2.5}{}$, where $F_\parallel=\SI{-0.84}{\femto\newton}$ and $v_\parallel=\SI{-0.098}{\micro\metre\,\second^{-1}}$. 
In the further course of the curves, the values rise until $\chi=3$ and then decrease again until $\chi=\SI{4}{}$, where the values saturate at $F_\parallel=\SI{-0.73}{\femto\newton}$ and $v_\parallel=\SI{-0.082}{\micro\metre\,\second^{-1}}$.

According to amount, the largest velocity $v_\parallel=\SI{-0.098}{\micro\metre\,\second^{-1}}$, found here for $\chi=2.5$, is about $80\%$ larger than the velocity of cone-shaped particles with aspect ratio $\chi=\SI{0.5}{}$ studied in previous work \cite{VossW2020}. 
Since the energy density $E=\SI{0.0227}{\joule\,\metre^{-3}}$ used in the present work is strongly smaller than the largest energy density $E_\mathrm{max}=\SI{4.9}{\joule \metre^{-3}}$ allowed by the U.S.\ Food and Drug Administration for diagnostic applications in the human body \cite{QBarnettEtAl2000}, the results can be extrapolated to higher energies.
To be able to do this, we need to make some assumptions. In experiments, the dependence of the velocity was measured to be proportional to the squared amplitude of the driving voltage \cite{AhmedBJPDN2016}, which is the same scaling as for the energy density \cite{Bruus2012}. Therefore, we can assume that the velocity is proportional to the energy density. 
This assumption leads to a factor $E_\mathrm{max}/E=216$. Rescaling the highest magnitude of the velocity, which is attained at $\chi=2.5$, to the higher energy density $E_\mathrm{max}$ results in $v_\mathrm{rescaled}=\SI{21.2}{\micro\metre\,\second^{-1}}$. For $\chi=2.5$, the values of the parameters describing the particle size are $\sigma=\SI{0.45}{\micro\metre}$ and $h=\SI{1.12}{\micro\metre}$. With the higher energy density the particle would thus move with a speed of roughly $\SI{19}{}$ body lengths per second. For the largest positive velocity, which corresponds to $\chi=\SI{1}{}$, the rescaled speed has the value $v_\mathrm{rescaled}=\SI{18.6}{\micro\metre\,\second^{-1}}$. Since the size parameters are now $\sigma=\SI{0.71}{\micro\metre}$ and $h=\SI{0.71}{\micro\metre}$, this speed equals $\SI{26.2}{}$ body lengths per second. Depending on the particular application, one can therefore choose an aspect ratio of $\chi=2.5$ or $\chi=1$ to reach a maximal absolute speed or a maximal speed related to the particle length, respectively. 
For medical applications, compact particles in a certain size range are necessary \cite{DavisCS2008,LuoFWG2018,ReinisovaHP2019} such that they do not block the blood flow. Therefore, the aspect ratio $\chi=1$ could be preferable for these types of applications.

The dependence of $F_{\parallel,p}$ and $F_{\parallel,v}$ on the aspect ratio is simpler. Both values keep their sign with $F_{\parallel,p}$ as a negative force and $F_{\parallel,v}$ as a positive force. The magnitude of both values increases until $\chi=1$ fast to $F_{\parallel,p} = \SI{-10.56}{\femto\newton}$ and $F_{\parallel,v} = \SI{11.32}{\femto\newton}$. Afterwards, the magnitude oscillates a little bit but with the tendency to increase slowly towards $F_{\parallel,p} = \SI{-13.60}{\femto\newton}$ and $F_{\parallel,v} = \SI{12.87}{\femto\newton}$ at $\chi=4$.

We now consider the propulsion parallel to the direction of propagation of the ultrasound wave. 
The perpendicular propulsion force $F_\perp$ increases for increasing $\chi$ slowly until $\chi=\SI{1.5}{}$ to $F_\perp=\SI{0.25}{f\newton}$. Then it is constant until $\chi=\SI{2.5}{}$. From $\chi=\SI{2.5}{}$ to $\chi=\SI{3}{}$ a strong increase occurs to $F_\perp=\SI{0.41}{f\newton}$. Subsequently, a slow further increase follows. 
The perpendicular velocity $v_\perp$ can be seen as roughly constant with $v_\perp=\SI{0.023}{\micro\metre\,\second^{-1}}$ until $\chi=\SI{2.5}{}$ and then it has a small rise to $\chi=\SI{3}{}$ with $v_\perp=\SI{0.03}{\micro\metre\,\second^{-1}}$. Afterwards, it is roughly constant again. This behavior can be understood as follows: 
In the direction of ultrasound propagation, two opposing forces act on a particle. These are the acoustic radiation force and the acoustic streaming force. Typically, for a particle with a size of about a micrometer, the acoustic radiation force is the dominant one \cite{BarnkobALB2012,MullerBJB2012,WiklundRO2012}. The scaling behavior of the acoustic radiation force is nontrivial for nonspherical shapes, but for a sphere it scales linearly with the particle volume \cite{Bruus2012}. Since we kept the particle volume constant, it is therefore reasonable that the velocity $v_\perp$ shows no strong change when the aspect ratio of the particle is varied. 

The value of the pressure component $F_{\perp,p}$ of the force $F_\perp$ increases for increasing $\chi$ and the value of the viscous component $F_{\perp,v}$ decreases roughly until $\chi=\SI{2.5}{}$, except for a slight intermediate growth of $F_{\perp,v}$ near $\chi=\SI{1.5}{}$. At $\chi=3$, there is a downwards oriented peak in the amplitude for both components. 
Afterwards, the amplitude increases again for both components, and from $\chi=\SI{3.5}{}$ onwards the values of both components increase slightly. 

The torque acting on these particle shapes is for all aspect ratios, according to amount, rather small. It decreases for increasing $\chi$ from $T=\SI{-0.005}{f\newton\,\micro\metre}$ to $T=\SI{-0.43}{f\newton\,\micro\metre}$, where the curve has a small local maximum at $\chi=\SI{2}{}$.
The corresponding angular velocity decreases from $\omega=\SI{-.004}{\second^{-1}}$ at $\chi=\SI{0.25}{}$ to $\omega=\SI{-0.033}{\second^{-1}}$ at $\chi=\SI{1.5}{}$, increases afterwards to $\omega=\SI{-0.023}{\second^{-1}}$ at $\chi=\SI{2.5}{}$, decreases again to $\omega=\SI{-0.06}{\second^{-1}}$ at $\chi=\SI{3}{}$, and remains there for larger values of $\chi$. 
The small values of the torques are negligible compared to the rotational Brownian motion of the particles and indicate that there is either no preferred orientation of the particles or a stable orientation is close to the orientation considered in the present work. 
Considering the aspect ratio $\chi=4$, where we found, according to amount, the largest torque $T=\SI{-0.43}{f\newton\,\micro\metre}$ and angular velocity $\omega=\SI{-0.06}{\second^{-1}}$, a change of the particle orientation by 90 degrees takes roughly $\SI{26}{\second}$. 
On the other hand, calculating the particle's diffusion tensor $\mathcal{D}=(k_\mathrm{B} T_0 / \nu_\mathrm{s}) \boldsymbol{\mathrm{H}}^{-1}$, where $k_\mathrm{B}$ is the Boltzmann constant, leads to a rotational diffusion coefficient $D_{33}=\SI{0.68}{\second^{-1}}$ of the particle, which implies that the particle orientation changes significantly by Brownian rotation on the time scale $\SI{1.46}{\second}$. This estimate clearly shows that the torques resulting from the ultrasound are so weak that they are dominated by Brownian rotation. 
However, if the angular velocity is rescaled in the same way as the translation velocity to an energy density $E_\mathrm{max}$, a change of the particle orientation by 90 degrees needs only $\SI{0.12}{\second}$ and is thus dominant compared to the Brownian rotation.

Concerning the components of the torque, the pressure component $T_p$ decreases (with superimposed fluctuations) for increasing $\chi$ from $T_p=\SI{0.02}{\femto\newton\,\micro\metre}$ to $T_p=\SI{-0.36}{\femto\newton\,\micro\metre}$, whereas the viscous component $T_v$ fluctuates (with stronger amplitude than for $T_p$) around zero. 

In summary, the force and velocity perpendicular to the direction of propagation of the ultrasound wave have a sign change at a particular value of the aspect ratio of the particle, the force and velocity parallel to the propagation direction do not change sign, and the torque is very weak.

Our results are in line with the available theoretical results on ultrasound-propelled particles from the literature. In the theory of \textit{Collis et al.\ }\cite{CollisCS2017}, the propulsion direction depends strongly on the acoustic Reynolds number $\beta=\rho_0 \sigma^2 \pi f / (2\nu_{\mathrm{s}})$, which ranges for our work between $\beta=\SI{0.2}{}$ for $\chi=4$ and $\beta=\SI{3.1}{}$ for $\chi=\SI{0.25}{}$. They found that particles can change their propulsion direction up to two times when increasing the acoustic Reynolds number, which is exactly what happens here. According to their theory, this should happen for $\beta \sim O(1)$, which is perfectly in line with the interval of values for $\beta$ we investigated.
A similar result was found experimentally \cite{AhmedWBGHM2016,SotoWGGGLKACW2016}. There, long cylinders with spherical caps at the ends corresponding to an aspect ratio $\chi=\SI{4.3}{}$ to $\chi=\SI{17.3}{}$ are moving towards their concave end \cite{AhmedWBGHM2016}, whereas the short half-sphere cups with aspect ratio $\chi=\SI{0.5}{}$ have a propulsion in the opposite direction \cite{SotoWGGGLKACW2016}. This is qualitatively the same behavior as for our particles, where the long ones with $\chi\gtrsim 1.5$ are moving in the direction opposite to their convex end and the shorter ones with $0.5\lesssim\chi\lesssim 1$ are moving towards the convex end.

\section{\label{conclusions}Conclusions}
We have studied the acoustic propulsion of nano- and microcones powered by a traveling ultrasound wave through direct numerical simulations. 
Our results show that the propulsion of the particles depends sensitively on their aspect ratio and includes both fast forward and fast backward motion. 
The strong dependence of the propulsion on the aspect ratio could be used to separate and sort artificial and natural cone-shaped particles (such as carbon nanocones) in an efficient and easy way with respect to their aspect ratio.
For later applications of cone-shaped ultrasound-propelled particles, e.g., in medicine, an aspect ratio of $\chi=1$ was identified as a very suitable choice, since it combines a compact particle shape with a large body-lengths-per-time speed.  
This finding also suggests to use cone-shaped particles with this aspect ratio as a more efficient particle design for future experiments. 

The obtained results are in good agreement with the literature and expand the understanding of acoustically propelled colloidal particles, which is helpful with regard to future experiments and applications in nanomedicine or materials science. 
Furthermore, the knowledge about the particle propulsion can be used to model this propulsion when describing the dynamics of the particles via Langevin equations \cite{WittkowskiL2012,tenHagenWTKBL2015} or field theories based on symmetry-based modeling \cite{MarenduzzoOY2007,TiribocchiWMC2015}, the interaction-expansion method \cite{BickmannW2020twoD,BickmannW2020b}, classical dynamical density functional theory \cite{WensinkL2008,teVrugtLW2020}, or other analytical approaches on time scales that are much larger than the period of the ultrasound. 

In the future, this study should be extended by considering other particle orientations and studying how the propulsion depends on the angle between the particle orientation and the direction of propagation of the ultrasound. Furthermore, the dependence of the propulsion on parameters like the ultrasound frequency, fluid viscosity, and pressure amplitude still need to be investigated.

\section*{Conflicts of interest}
There are no conflicts of interest to declare.

\begin{acknowledgments}
We thank Patrick Kurzeja for helpful discussions. 
R.W.\ is funded by the Deutsche Forschungsgemeinschaft (DFG, German Research Foundation) -- WI 4170/3-1. 
The simulations for this work were performed on the computer cluster PALMA II of the University of M\"unster. 
\end{acknowledgments}

\nocite{apsrev41Control}
\bibliographystyle{apsrev4-1}
\bibliography{control,refs}

\end{document}